\documentclass[lettersize,journal]{IEEEtran}
\usepackage{amsmath,amsfonts}
\usepackage{array}
\usepackage{textcomp}
\usepackage{stfloats}
\usepackage{url}
\usepackage{verbatim}
\usepackage{graphicx}
\usepackage{authblk}
\usepackage{subcaption}
\usepackage{cite}
\usepackage[ruled,vlined,linesnumbered]{algorithm2e}
\usepackage{booktabs}
\usepackage{hhline}
\usepackage{xcolor}

\usepackage{hyperref}
\newcolumntype{M}[1]{>{\centering\arraybackslash}m{#1}}
\renewcommand{\vec}[1]{\boldsymbol{#1}} 

\begin{document}

\title{Learning from Synthetic InSAR with Vision Transformers: The case of volcanic unrest detection}

\author{
Nikolaos Ioannis Bountos\IEEEauthorrefmark{1}\IEEEauthorrefmark{2}, Dimitrios Michail\IEEEauthorrefmark{2}, and Ioannis Papoutsis\IEEEauthorrefmark{1},~\IEEEmembership{Member,~IEEE}
\thanks{\IEEEauthorrefmark{1}Institute of Astronomy, Astrophysics, Space Applications \& Remote Sensing, National Observatory of Athens (e-mail: \{papoutsis, bountos\}@noa.gr)}
\thanks{\IEEEauthorrefmark{2}Department of Informatics \& Telematics, Harokopio University of Athens (e-mail: michail@hua.gr)}
}

\maketitle

\begin{abstract}
The detection of early signs of volcanic unrest preceding an eruption, in the form of ground deformation in Interferometric Synthetic Aperture Radar (InSAR) data is critical for assessing volcanic hazard.
In this work we treat this as a binary classification problem of InSAR images, and propose a novel deep learning methodology that exploits a rich source of synthetically generated interferograms to train quality classifiers that perform equally well in real interferograms.
The imbalanced nature of the problem, with orders of magnitude fewer positive samples, coupled with the lack of a curated database with labeled InSAR data, sets a challenging task for conventional deep learning architectures.  
We propose a new framework for domain adaptation, in which we learn class prototypes from synthetic data with vision transformers. We report detection accuracy that 
 amounts to the highest reported accuracy on a large test set for volcanic unrest detection.
Moreover, we built upon this knowledge by learning a new, non-linear, projection between the learnt representations and prototype space, using pseudo labels produced by our model from an unlabeled real InSAR dataset. This leads to the new state of the art with $97.1\%$ accuracy on our test set.
We demonstrate the robustness of our approach by training a simple ResNet-18 Convolutional Neural Network on the unlabeled real InSAR dataset with pseudo-labels generated from our top transformer-prototype model.
Our methodology provides a significant improvement in performance without the need of manually labeling any sample, opening the road for further exploitation of synthetic InSAR data in various remote sensing applications.
\end{abstract}

\begin{IEEEkeywords}
volcanic unrest, InSAR, vision transformer, prototype learning, domain adaptation.
\end{IEEEkeywords}

\section{Introduction}
\IEEEPARstart{T}{he} availability of Copernicus Sentinel-1 data, acquired with global coverage on a systematic basis, has spurred the development of new applications in Remote Sensing. The sheer volume of the data generated has allowed the use of scalable deep learning methods that are able to efficiently and accurately automate information extraction from these rich data archives{\color{blue} ~\cite{zhu2021deep}}, detect subtle signals hidden in satellite imagery{\color{blue} ~\cite{wang2019sar}}, and predict key environmental variables which in turn can be fed to physical models{\color{blue} ~\cite{ban2020near}}, e.g. for hazard and risk assessment{\color{blue} ~\cite{bianchi2020large, el2020reasoned}}.

Volcanic hazard in particular, is critical for disaster risk reduction, especially near urban and peri-urban areas, with more than 800 million people living within 100km from an active volcano, while 29 million within 10km~\cite{brown2017volcanic}, posing a valid threat to the population. Direct and severe economic loss is also a direct impact of a volcanic eruption. For example, the 2010 Eyjafjallaj\"{o}kull Icelandic volcano moderate eruption led to the closure of north and central European airspace, the cancellation of $\sim$100,000 flights and the stranding of $\sim$10 million passengers, which according to Oxford Economics~\cite{mazzocchi20102010} resulted to  a total global economic impact of $\sim$5bn €, as the major hubs of London, Amsterdam, Paris and Frankfurt were virtually shut down due to the effects of the ash clouds.

In fact, according to Siebert et al.~\cite{siebert2011volcanoes}, about 1,500 volcanoes are known to have erupted in the last 12,000 years and about 700 of these have erupted at least once in historical times. Currently, worldwide, about 100 volcanic unrests are observed yearly, and about half of them become observable eruptions~\cite{capes2012international}. However, it is estimated that less than 10\% of active volcanoes are monitored on a systematic basis~\cite{loughlin2015global}, and therefore volcanic hazards are monitored occasionally, or not monitored at all from  global volcano observatories, such as the Geohazard Supersites and Natural Laboratories initiative~\cite{salvi2016geo}, that has a mandate to assess geohazards globally for critical areas, in support of Disaster Risk Reduction.

Detecting early signs of volcanic activity can be of paramount importance to promptly mobilise scientific teams, deploy sensing equipment on the ground and alert civil protection authorities. Interferometric Synthetic Aperture Radar (InSAR) products are a rich information source that is used to detect ground deformation associated with volcanic unrest~\cite{doi:https://doi.org/10.1029/GM116p0207}. Such deformation is statistically linked to an eruption~\cite{biggs2014global} and can be detected prior to the event~\cite{FURTNEY201838}. The deformation appears in the wrapped InSAR data in the form of interferometric fringes. Unfortunately, the different atmospheric conditions between the two temporally separated SAR acquisitions that are needed to form an interferogram, can give rise to similar fringe patterns. The main types of atmospheric artefacts are vertical stratification that is correlated with topography and is caused by changes of the refractive index of the troposphere~\cite{hanssen2001radar}, and turbulent mixing and vapors caused by liquid and solid particles of the atmosphere. A thorough analysis on the impact of the coincident nature of deformation and atmosphere at (Strato)volcanoes, especially in the presence of strong topography, can be found in the work of Parker et al.~\cite{PARKER2015102}. Therefore, the task of automatically detecting interferograms with fringes attributed to ground deformation is challenging.

\IEEEpubidadjcol
Recent works, motivated by the advances of computer vision, have attempted to employ convolutional neural networks (CNN) to solve the task of deformation detection in single wrapped and/or unwrapped InSAR data. Different CNN architectures have been tested. e.g. AlexNet~\cite{anantrasirichai2019deep,anantrasirichai2018application} and VGG-based~\cite{gaddesasimultaneous}, while custom CNN models have also been designed~\cite{valade2019towards}.
A CNN workflow has also been used to detect fringes associated with an earthquake \cite{brengman2021identification}. The work of Gaddes et~al.~\cite{gaddes2019using} is the first effort that considers InSAR time-series approaches to automatically detect volcanic deformation, using Independent Component Analysis and clustering on synthetic unwrapped interferograms to retrieve signals of geophysical significance. Albino et al.~\cite{https://doi.org/10.1029/2019JB017908} adopt the cumulative sum control chart, an anomaly detection approach for monitoring small changes in unwrapped InSAR time-series. Anantrasirichai et al.~\cite{anantrasirichai2019application} propose a hybrid method that blends time-series methods with deep learning. The authors use singular value decomposition on short duration interferograms to obtain the time series of cumulative ground deformation, which they then wrap and classify using a pre-trained CNN model. This method is especially fit for detecting slowly evolving unrest episodes. Finally, Sun et al~\cite{sun2020automatic} address the problem as a denoising, regression one, and use a modified U-Net architecture on synthetic and augmented unwrapped time-series data.  

Our focus is on the volcanic unrest classification problem for a single wrapped interferogram, aiming for simplicity, robustness, rapid classification response, and scalability potential towards global coverage. We avoid using unwrapped interferograms, since phase unwrapping is prone to unwrapping errors. Similarly, we do not exploit time-series data to limit processing needs, such as inversion~\cite{anantrasirichai2019application}, and storage requirements. For single wrapped interferogram classification deep CNNs seem to prevail in literature.

However, even though CNN methods have proved to be robust and applicable in multiple domains such as medicine \cite{xu2015deep}, remote sensing \cite{papoutsis2021efficient}, self-driving cars \cite{chen2017end}, biology\cite{shujaat2020pcpromoter} and others, they tend to be exceptionally data-hungry. Deep CNN architectures require large amounts of curated labeled data, however there is no such database for InSAR data. In addition, our problem is inherently imbalanced, since the number of negative samples with no volcanic deformation fringes, surpass the set of positive samples with volcanic deformation fringes, by orders of magnitude. To mitigate these problems, researchers have exploited pretrained models from optical datasets such as ImageNet \cite{deng2009imagenet}, data augmentation techniques and synthetic data to address the scarcity of labeled InSAR data \cite{marmanis2015deep, bar2015chest,anantrasirichai2019deep,anantrasirichai2018application,gaddesasimultaneous,valade2019towards}. However, Bountos et~al.~\cite{bountos2021self} showed that training a model with data from a different domain, will incur a drop in classification performance when compared to models trained in the same domain. In fact, the work of Bountos et~al.~\cite{bountos2021self} is, to the best of our knowledge, the state-of-the-art for volcanic unrest detection with 91\% classification accuracy, by developing a  self-supervised pretraining strategy based on unlabeled InSAR data.

In this work we exploit synthetically generated InSAR patches, which we can generate in abundance, and design a training framework based on learning representative class prototypes with vision transformers. Our method is able to generalize well to the real InSAR data domain, without the need for human annotations. To summarize, our main contributions are: 
\begin{itemize}
    \item We successfully train both CNNs and visual transformers for the binary volcanic unrest detection problem using only
          synthetically generated InSAR data.
    \item We present a novel prototype learning scheme using vision transformers, which significantly surpasses the current state of the art.
    \item We successfully transfer knowledge from synthetically generated InSAR to real data by designing an unsupervised self-labeling domain
          adaptation pipeline.
    \item We provide a framework for robust InSAR data pseudo-labeling and demonstrate its effectiveness.
    \item We publish all trained models and code used in this work\footnote{https://github.com/Orion-AI-Lab/PrototypeInSAR}.
\end{itemize}

\section{Related Work}
\subsection{Domain Adaptation}
Given the lack of labeled InSAR datasets, there are previous works that use synthetic InSAR data to train their models. Brengman et~al.~\cite{brengman2021identification} used synthetic InSAR to classify co-seismic deformation, proposing a CNN architecture called SarNet, and evaluated its performance on real data. The authors report an accuracy of $47.62\%$ on a test set of 32 real InSAR, hence, in order to enhance the generalisation capability of their model, they choose to finetune it by injecting real augmented InSAR data in the learning process. The reported accuracy ranges between 59.22\% and 85.22\%.

Gaddes et~al.~\cite{gaddesasimultaneous} use the first five convolutional blocks of a pretrained VGG16\cite{simonyan2014very} to get an image representation on synthically generated InSAR data over volcanic areas, and then feed it to a Multi Layer Perceptron (MLP) for the classification task. Their classes include dyke, sill and no deformation. To evaluate their method in the real domain, they generate a set of 52 interferograms and evaluate their model achieving an accuracy of 65\%. To improve their results, they finetune on a set of 173 real interferograms, reporting an updated accuracy of 67\% for the dyke (3 samples), 82\% for Sill (17 samples) and 84\% (32 samples) for the no deformation patches, which translates to $\approx82.3\%$ accuracy for the binary classification task. 

Anantrasirichai et~al.~\cite{anantrasirichai2019deep} generate synthetic data in order to train an AlexNet for binary classification. They report $41/363$ true positives and $0$ false negatives. After retraining the network with a combination of real and synthetic data the reported results improve to $41/52$ true positives and 1 false negative.  Valade et al.~\cite{valade2019towards} also generate synthetic InSAR data to train a custom CNN model that outputs the associated phase gradients and phase decorrelation masks. In production, the authors report high classification accuracy on 4 positive and 256 negative real interferograms for two volcanoes. 

Considering time-series approaches, in~\cite{sun2020automatic} Sun et al. generate synthetic data assuming Mogi model as the deformation source and produce 20,000 time-series groups, where each group contains 20 time-consecutive pairs of unwrapped surface displacement maps. Gaddes et al.,~\cite{gaddes2019using} use synthetic data to train their non-deep learning model in an unsupervised approach, and test their approach in Sierra Negra volcano using unwrapped interferograms. Finally, in ~\cite{anantrasirichai2019application} the authors use synthetic wrapped InSAR to pre-train a CNN model that is then used as a module for a more complex chain that is able to detect subtle deformation signals.

All of the above methods encounter the problem of the covariate shift which affects the generalization to the real domain. The covariate shift is caused by a change in the distribution of the input samples while the label distribution remains intact. This issue has been studied extensively by the computer vision community.
Since we assume no labels from the real domain we focus on the case of unsupervised domain adaptation. In \cite{coral} Sun et~al. propose a simple framework where they align the second order statistics of the features of source and target domains.  CORrelation ALignment (CORAL) was combined with deep neural networks in \cite{dcoral} where they extended the initial method introducing the coral loss that minimizes the distance between the second order statistics of the source and target domains in deep layers, complementing the standard classification loss. In a similar manner, Tzeng et~al.~\cite{tzeng2014deep} proposed to minimize the Maximum Mean Discrepancy\cite{borgwardt2006integrating} of deep representations along with the classification loss. 
In \cite{pmlr-v37-ganin15} the authors propose to incorporate a gradient reversal layer along with a domain classifier in order to learn domain invariant features while minimizing a standard classification loss. The gradient reversal layer multiplies the gradient flowing from the domain classifier with a constant negative factor penalizing the feature extractor for the discriminator's good performance. This "game" results in a feature extractor that can produce domain invariant representations with good classification potential. 

Another set of approaches focuses on the style transfer between source and target domain. In that direction Zhu et~al.~\cite{zhu2017unpaired} propose a framework that is able to transform images from domain A to domain B and vice versa. Building upon that, Murez et~al.~\cite{murez2018image} propose a methodology for image to image domain adaptation. They enforce the learnt representations to be reconstructable for both domains, while making the latent space domain agnostic via a domain classifier. Furthermore they include a translation loss where an image from the source domain is encoded and then reconstructed as an image from the target domain and vice versa. The translation loss comes from a domain classifier that attempts to identify the domain of the image. Furthermore, they apply a cycle consistency loss where they try to regenerate the original image from its fake opposite domain counterpart. Finally, they integrate a classification loss for the target domain, where source samples are translated to the target retaining their labels. This enables the target domain to be included in the supervised training.

Following a completely different line, self-supervised learning has proved to be very effective on the task of domain adaptation and dealing with label scarcity. Self-supervised methods are able to exploit the wealth of the information hidden inside unlabeled data to learn features without any supervision that can generalize well for different classification tasks.
Thota et. al~\cite{thota2021contrastive} used a contrastive learning framework to learn domain invariant features for fully unsupervised domain adaptation. Carlucci et. al~\cite{carlucci2019domain} and Bucci et. al~\cite{bucci2021self} utilized classic self-supervised learning methods e.g Jigsaw Puzzles~\cite{noroozi2016unsupervised} and rotation recognition on multiple domains to create robust, multi-domain representations while concurrently addressing the task of object classification in a supervised manner. Furthermore, Bucci et.al~\cite{bucci2019tackling} tackled the problem of partial domain adaptation where the target contains a subset of the source domain's class set using the same idea of exploiting a Jigsaw puzzle as a side task that can reduce the domain gap. Finally, in \cite{bucci2020effectiveness} the authors utilized the self-supervised task of predicting image rotation to address the task of open set domain adaptation where the target domain's class set is not restricted to the source's class set.
Our work is directed towards the creation of a low dimensional space where samples from both source and target domains revolve around the respective class prototypes. These prototypes are learnable and jointly trained along with the encoder.

\subsection{Prototype Learning}
Prototype learning methods have been traditionally used in machine learning for computer vision applications. Typical prototype based methods include the standard k-Nearest Neighbors algorithm (k-NN) and the learning vector quantization algorithm \cite{kohonen1990self}. However, in modern computer vision there have been only a few works related to prototypes. Yang et~al.~\cite{yang2018robust} introduce the general convolutional prototype learning framework that combines the standard  convolutional neural network based approaches with learnable prototypes for image classification. In that setting the class of a sample is assigned to the class of its nearest prototype.
Li et~al.~\cite{li2020prototypical} uses prototypes in the setting of self-supervised contrastive learning in order to find embeddings that are able to encode the semantic structure of the data, enforcing samples to have more similar embeddings with their respective prototypes compared to the rest of the prototypes. Moreover, Huang et~al.~\cite{huang2020relational} combined graph neural networks with prototype learning for the task of action localization in videos.

Pinhero et.al \cite{pinheiro2018unsupervised} address a similar problem to ours for computer vision applications, combining domain adaptation with prototypes. The authors use a domain classifier to discriminate between source and target samples along with a gradient reversal layer, while performing the classification based on similarity with class prototypes. In contrast to our approach, where the prototypes are learnable, in Pinhero et.al~\cite{pinheiro2018unsupervised} they are defined as the average representation of the source dataset. Furthermore, in our case the classification is based on the distance of samples to the respective prototypes in a space obtained by projecting the learnt representations on a low dimensional prototype space. On the contrary, Pinheiro et.al use a a bilinear operation as a similarity measure on which the classification is based.

Prototype learning approaches have also been used in remote sensing. Hua et~al.~\cite{hua2021aerial} use a prototype learning approach to recognize multiple aerial scenes in an image. They learn prototypes from different known scenes and when a new query image is fed into the network they retrieve the relevant scenes by using a multi-headed attention like mechanism, where the query is the new image and the keys are the known scenes. Then, they combine the results of the different heads and feed it into a final fully connected classification layer. Finally, Zhang et~al.~\cite{zhang2020global} learn class related prototypes on a feature space obtained from an encoder for the task of hyperspectral image classification. The classification is handled with a nearest neighbor approach. According to the authors this method performed better than other approaches in a limited labeled data regime.

\subsection{Vision Transformers}
Transformers have taken the natural language processing (NLP) community by a storm \cite{vaswani2017attention, devlin2018bert}. This success motivated the computer vision community to adapt it to its needs \cite{dosovitskiy2020image, touvron2021training,arnab2021vivit}. The resulting vision transformers, which from now on we will refer to as ViT, operate by splitting the image into a sequence of patches that are fed to a standard transformer encoder. Given the lack of inductive bias in comparison to CNN's, training a ViT from scratch is quite challenging. The concept of locality is completely absent even in the presence of positional embeddings since initially they contain no useful information\cite{dosovitskiy2020image} making it extremely difficult to train properly in small datasets, compared to CNNs. Recent works have tried to work around this issue by investigating sophisticated methods to train ViTs\cite{touvron2021training} or combining them with the inductive bias of CNNs~\cite{d2021convit, liu2021swin}. A standard approach to properly utilize ViTs is to pretrain them in large datasets and then apply them to downstream tasks \cite{dosovitskiy2020image, steiner2021train}. We follow this approach throughout this work.

It is not the first time that vision transformers are exploited for remote sensing tasks. Bazi et al.~\cite{rs13030516} and Papoutsis et al.~\cite{papoutsis2021efficient} use them for the task of satellite scene classification, while Horv\`ath~\cite{horvath2021manipulation} address the satellite image manipulation detection problem, in which the authors try to detect satellite images that have been tampered. However, there has been no prior work on synthetic to real satellite data adaptation, let alone the InSAR domain.

\section{Approach}
\begin{figure*}[ht!]
    \centering
    \includegraphics[width=0.9\textwidth]{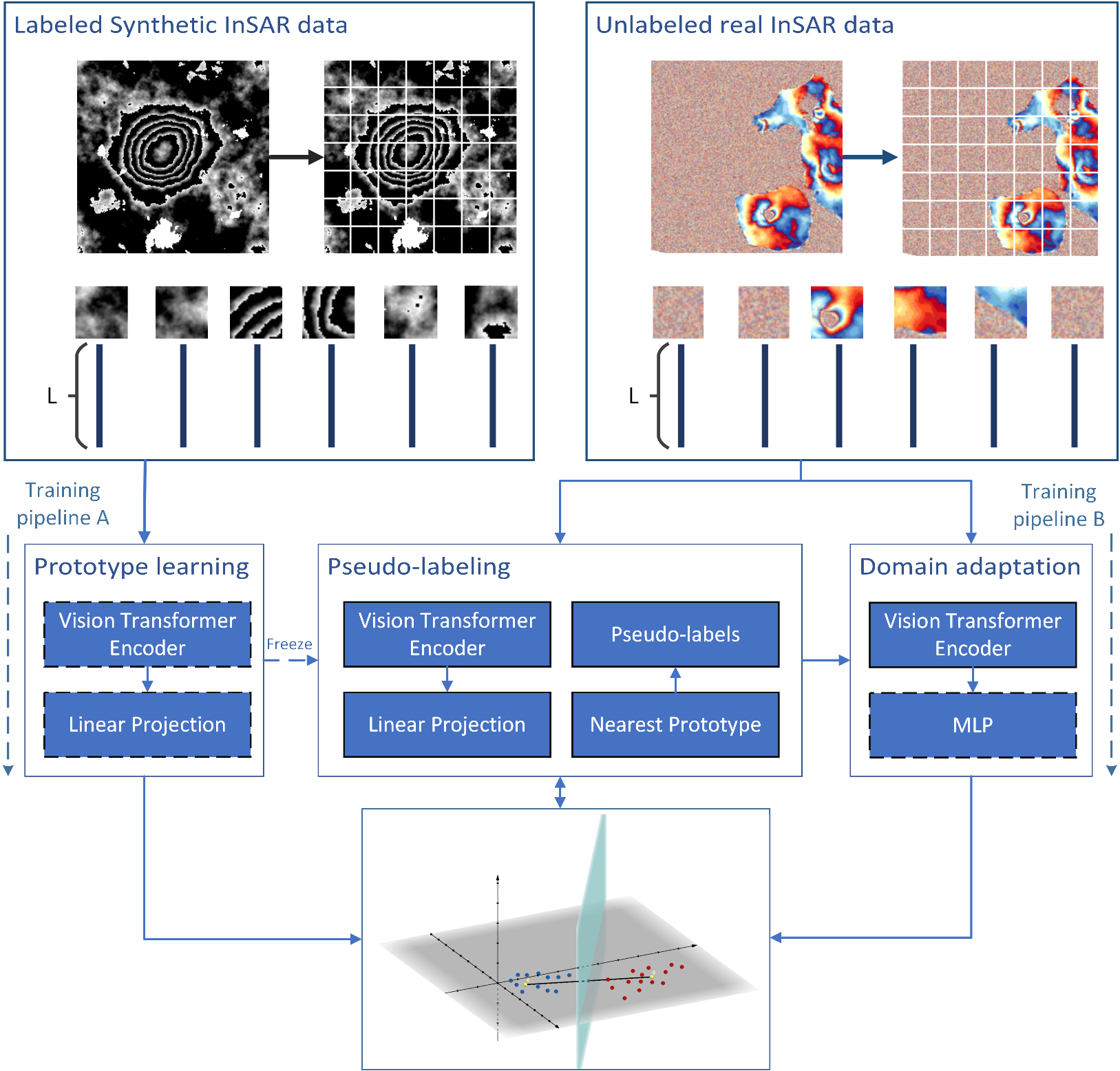}
    \caption{This figure depicts the full training procedure. At training time, a synthetic InSAR sample is split in patches and fed to the Transformer Encoder. The output of the encoder is then projected to the prototype space using a learnable linear projection. The prototypes are depicted in the graph with yellow, whereas the two classes with blue and red. The samples are assigned to the class with the closest prototype. At the pseudo training stage, the trained model generates pseudo labels on an unlabeled real InSAR dataset. Then the linear projection module is replaced with a more complex, non-linear MLP for domain adaptation. All parts of the initial model are frozen including the prototypes leaving the MLP the only trainable module. The resulting model retains its classification abilities on the synthetic data while improving on the real domain. In this figure, dashed line boxes correspond to trainable modules, whereas solid line boxes correspond to non-trainable modules.}
    \label{fig:domain}
\end{figure*}

In this section we discuss the main components of our architecture. 
We start by defining a Swin Transformer as the backbone of the architecture, adapting it to the needs of a prototype learning approach. Then we introduce a domain adaptation component to further improve the model's generalization to the real domain. The full pipeline is presented both schematically (Figure~\ref{fig:domain}) and in pseudocode (Algorithm~\ref{alg:iterative}). 

\subsection{Transformer-based encoder}\label{sec:encoder}

We utilize a vision transformer as our feature extractor $f(x)$. We focus on the Swin Transformer but since our framework 
is encoder invariant, we also report results using other architectures.
Swin Transformer \cite{liu2021swin} builds upon ViT in order to create a general purpose efficient backbone for computer vision tasks. In contrast with standard ViT, Swin Transformer aims to learn hierarchical features while calculating self-attention in a more efficient manner. It deviates from traditional vision transformers by injecting inductive bias, locality and hierarchy into the transformer architecture \cite{xie2021self}. As in standard ViTs, the image is split into a sequence of patches that constitute the input to the Transformer Encoder (see Figure~\ref{fig:input_sample} for an example image split into patches). Self-attention is computed in local windows in contrast with standard ViTs where self-attention is computed globally. A window is defined as a set of $M \times M$ patches, where $M$ is fixed e.g. $M=7$ (default value). The fixed size of windows reduces the quadratic complexity to the image size of typical self-attention computation in ViTs to linear complexity to the number of patches for the Swin Transformer. In order to allow communication between windows, a shifting windows approach is used between consecutive transformer blocks. In this case, the windows of the preceding layer are shifted, creating windows that connect previously disconnected patches. Furthermore, after each set of swin transformer blocks, a patch merging layer is applied to fulfill the promise of hierarchical features. For a more detailed view of the architecture, we refer the reader to~\cite{liu2021swin}.

\begin{algorithm}[th!]
    \SetKwInput{KwInput}{Input}
    \SetKwInput{KwOutput}{Output}
    \SetAlgoLined

    \KwInput{synthetic labeled data $S = \{(x_i,y_i)\}^n_{i=1}$}
    \KwInput{real unlabeled data $A = \{(x_i)\}^{a'}_{i=1}$}
    \KwInput{training epochs $epochs_S$ on synthetic data }
    \KwInput{training epochs $epochs_P$ on pseudo-labeled data }
    \KwInput{$with\textnormal{-}pseudo\textnormal{-}training$ flag}

    \KwOutput{trained model $model$}

    initialize the parameters $\theta$ of the feature extractor $f(x)$\;
    initialize prototypes $M$\;
    let $model$ be the prototype based model\;
    \For{epoch in range($epochs_S$)}{
     \For{each batch $\vec{b_s} = \{(x_i, y_i)\}^B_{i=1} \in S$}{
         $output, minus\_distance, features \leftarrow  model(\vec{b_s})$\;
         $prediction = argmax(minus\_distance)$\;
         $l_p = l(\vec{output};\theta;M) + \lambda        pl(\vec{features},prediction;\theta;M)$\;
         adjust $\theta$ and $M$ to optimize $l_{p}$\;
         }
    }
    
	\If{$with\textnormal{-}pseudo\textnormal{-}training$}{
	generate pseudo-labels for $A$\;
        freeze model and prototypes\;
        $model.fc = MLP()$\;
        initialize the parameters $\theta'$ of the $MLP$\;
        \For{epoch in range($epochs_P$)}{
            \For{each batch $\vec{b_a} = \{(x_i, y_i)\}^B_{i=1} \in A$}{
                $output, minus\_distance, features \leftarrow  model(\vec{b_a})$\;
                $prediction = argmax(minus\_distance)$\;
                $loss = l(\vec{b_a};\theta, \theta';M) + pl(\vec{features},prediction;\theta,\theta';M)$\;
                adjust $\theta'$ to optimize $loss$\;
            }
        }
            
       }
    \Return{parameters $\theta$ and prototypes $M$}\;
    \caption{Pseudo-code for the training pipeline.}
    \label{alg:iterative}
\end{algorithm}

\subsection{Prototype Learning}\label{sec:prototype-learning}

Instead of classifying using a softmax layer, we feed the encoder's output to a prototype learning module which 
learns prototypes of each class during training. We build upon the
Convolutional Prototype Learning framework introduced in \cite{yang2018robust}. In our case the transformer-based encoder is used to extract features.
On top of the features we learn prototypes of each class. The classes are assigned via a nearest neighbors approach of the sample's representation with
the class prototypes.
Besides the classification loss we add a prototype loss (PL). The PL loss has the tendency to draw samples from the same class closer together and samples
from different classes further away. As discussed in Section~\ref{sec:experiments}, this results in more robust representations. Furthermore, due to the very nature of the
prototype loss, the resulting latent space is more k-means friendly.

Given the binary nature of the problem we use two prototypes, one for each class $m_i$ for $i \in \{0, 1\}$. This can be extended to any number of classes and prototypes for each class. In particular $m_{i,j}$ for $i \in \{1,...,C\} $ and $ j \in \{1,...,K\}$, where $C$ is the number of classes and $K$ the number of prototypes for each class. These prototypes are learnt during training time. The classification is based on a nearest prototype approach using the Euclidean distance. We experiment 
on different dimensionalities for the prototype space. 

We use the distance based cross entropy loss function (DCE) as defined in \cite{yang2018robust}. Given a distance function $d(f(x),m_{i,j}) =  \left\lVert f(x) - m_{i,j} \right\rVert_{2}^2$ we define the probability of a sample $(x,y)$ to belong to prototype $m_{ij}$ as:
\begin{equation}
    p(x\in m_{ij}|x)= \frac{e^{-\gamma d(f(x),m_{ij})}}{\sum_{k=1}^{C}\sum_{l=1}^Ke^{-\gamma d(f(x),m_{kl})}},  
\end{equation} 
where $\gamma$ is hyper-parameter that controls the hardness of probability assignment.
We can then define the probability of a sample $x$ belonging to class $y$ as:
\begin{equation}
    p(y|x) = \sum_{j=1}^Kp(x\in m_{yj}|x) .
\end{equation}
The distance based cross entropy loss is then calculated as: 
\begin{equation}
    l((x,y);\theta;M)= -log(p(y|x)), 
\end{equation} where $\theta$ are the parameters of the feature extractor and $M$ the set of prototypes.

Finally we add as a regularizing term a Prototype Loss that attempts to minimize the distance of the representation with the
closest prototype of the correct class 
\begin{equation}
    pl((x,y),\theta,M) = \left\lVert f(x) - m_{yj}\right\rVert_{2}^2 .
\end{equation}
The combined loss is then defined as
\begin{equation}
l((x,y);\theta;M) + \lambda pl((x,y);\theta;M),    
\end{equation}
where $\lambda>0$ controls the effect of the extra prototype loss.

As stated by Yang~et~al.~\cite{yang2018robust}, the PL loss has the ability to pull samples closer to their respective prototypes making the representations of samples from the same class compact and increasing the distance between different classes. Additionally, the classification loss tends to boost the class separation property of the representations. Thus, by combining them we can learn intra-class compact and inter-class separable representations. For low-dimensional prototypes we let both loss terms contribute equally. When we increase the prototype dimension we downscale the PL-loss accordingly. We denote this architecture when combined with an 
encoder X as X-PL.

\subsection{Domain Adaptation Projection Training}
\label{sec:pseudo}

An additional boost in performance can be achieved by introducing a supplementary domain adaptation component. 
The idea is to utilize self-labeling~\cite{triguero2015self} in an effort to improve the model's adaptation
to a new domain in an unsupervised setting. In self-labeling domain adaptation approaches, it is common to use 
the current model trained under a source domain in order to assign pseudo-labels for all samples to a target domain. 
These pseudo-labels are then considered as ground-truth and the model is retrained.

In our case we use an unlabeled, real InSAR dataset and proceed with the pseudo-label generation. Our assumption is that our encoder is generic 
enough to produce good representations in the real domain too. This is validated in our experiments (see Section~\ref{sec:experiments}) and the visualization of the prototype space in Figure~\ref{fig:protospace}. Given a good encoder and representative class prototypes, what remains is the projection to the prototype space. Instead of using the projection learnt for synthetic data, we opt to learn a new, non-linear projection tuned from real InSAR patches. We thus, freeze both our encoder and the learnt prototypes and throw away the final layer which projects the encoder's representations to the prototype space. This last layer is replaced with a $3 \times$ Layer MLP which is trained using the produced pseudo-labels. Besides improving performance on the real domain, the new projection retains the good embeddings learnt for the synthetic domain.

\section{Experiments}\label{sec:experiments}

\begin{figure}[ht]
    \centering
    \includegraphics[width=0.49\textwidth]{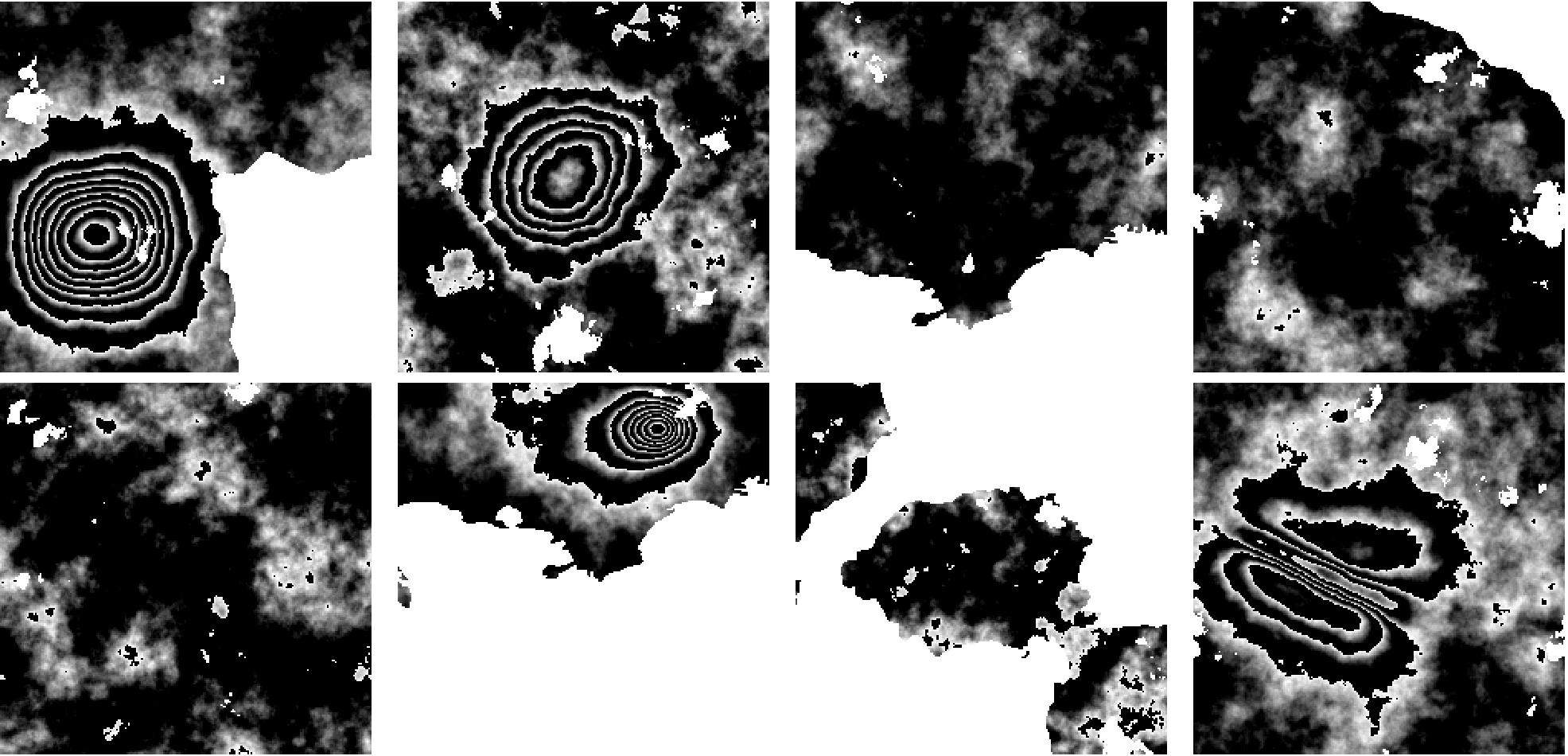}
    ~\\
    \includegraphics[width=0.49\textwidth]{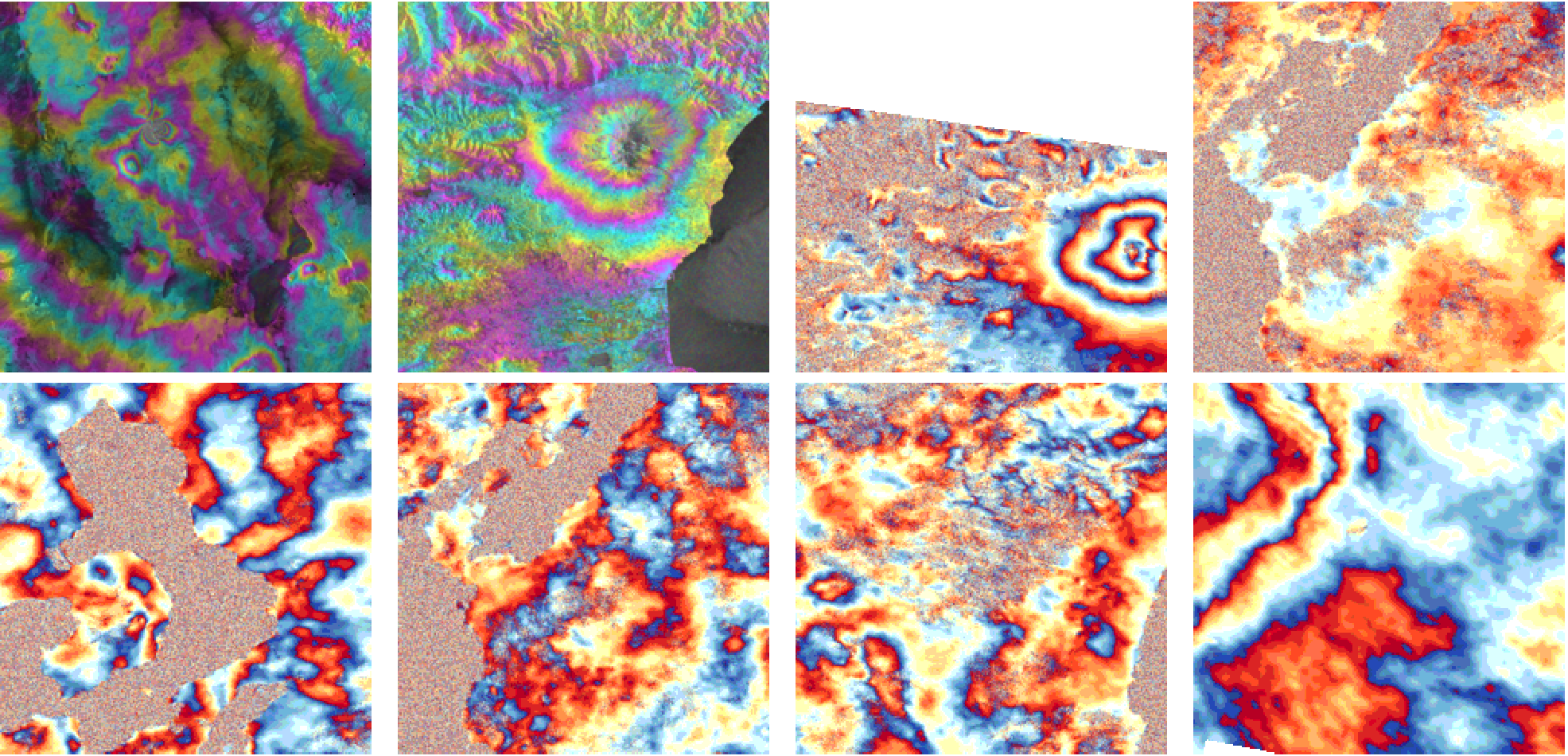}
	\caption{Samples drawn from the synthetic dataset (top) and from the real test 
	dataset (bottom).}
    \label{fig:all_samples}
\end{figure}

In this section, we investigate how well we can transfer knowledge learnt on synthetic data to the real domain. 
Our training set consists of $25,000$ synthetically generated InSAR samples created using the \href{https://github.com/matthew-gaddes/SyInterferoPy}{SyInterferoPy}~\cite{https://doi.org/10.1029/2019JB017519} generator, which produces random interferograms over a collection  of  subaerial  volcanoes. The interferograms arise from the synthesis of i) deformation signals considering simple dyke, sill, or Mogi sources, ii) a topographically correlated atmospheric phase screen (APS), iii) a turbulent APS, iv) phase gradients, and v) the superposition of regions of incoherence. Interferograms without deformation can also be synthesised. In our synthetic dataset, deformation is in the range [10cm 25cm], turbulent APS has on average a maximum strength of 2cm and is correlated on a 5km scale. We also vary the rad/km of delay for the topographically correlated APS. For the Mogi volcanic source the random variables are the expected magma chamber volume change and the source depth, for the sill we vary parameters like strike, depth, width, length and dip, while for the dyke we vary in addition the depth of its top segment. As seen in Figure~\ref{fig:all_samples}, the synthesized interferograms are a very good, but simplified approximation of reality. Interferometric fringe patterns in particular appear in synthetic data to be clearer with respect to real InSAR data. This implies that domain adaptation would be needed to improve classification performance.

For the validation set we construct a second synthetic dataset containing 3,361 deformation and 5,000 non-deformation patches. To evaluate and compare
the performance of our models with the current state-of-the-art, we use as test set the C1 dataset, introduced in \cite{bountos2021self}, containing real InSAR patches from Comet-LiCS portal~\cite{lazecky2020licsar,morishita2020licsbas,wright2016licsar,lawrence2013storing}, covering several volcanoes such as Taal, Cerro Azul, Sierra Negra, Fagradalsfjall, Etna, La Cumber and Erta Ale. More information about the volcanoes and their recent activity can be seen in Table~\ref{tab:c1_details}.
The details of each dataset, such as the number of positive and negative samples, can be seen in Table~\ref{tab:data}. Finally, our experiments were conducted using a NVIDIA GeForce RTX 3090 gpu.

\begin{table*}[]
\centering
\caption{Information on the volcanoes included in the C1 dataset and their recent activity.}\label{tab:c1_details}
\begin{tabular}{lllll}
 Volcano & Location  & Volcano Type & Recent Eruption & Deformation Model  \\ 
 \toprule
 Etna& Italy & Stratovolcano &  December 2018 &Dyke \\ \hline
 Taal& Philippines &Caldera  & January 2020 & Dyke   \\ \hline
 Fagradalsfjall& Iceland & Tuya & March 2021 &  Dyke  \\ \hline
 Erta Ale & Ethiopia & Shield & January 2017  & Dyke   \\ \hline
Cerro Azul & Isabela Island & Shield & - &   Dyke  \\ \hline
Sierra Negra  & Isabela Island & Shield & June 2018  &  Sill\\ \hline
La Cumber& Fernandina&Shield & September 2017, January 2020 &Dyke \\
\bottomrule
\end{tabular}

\end{table*}

\begin{table}[ht]
\centering
    \caption{Dataset break down.} \label{tab:data}
    \begin{tabular}{cccc}
    \toprule
    Data Source & Positive & Negative & Purpose\\ \hline
    Synthetic & 17976 & 7024 & Train\\ \hline
    Synthetic Validation & 3361& 5000 & Validation\\ \hline
    C1 & 404&365& Test\\ \hline
    \bottomrule
    \end{tabular}
\end{table}

\subsection{Investigating the encoder architecture} \label{sec:best-encoder}

Our first experiments investigate different encoder architectures. We evaluate 
the performance of standard convolutional neural network architectures such as ResNet18~\cite{he2016deep}, VGG16 and DenseNet121~\cite{huang2017densely} 
and more recent vision transformer architectures such as ConvViT, DeiT and Swin. In all cases we use a standard 
softmax layer for the classification. We train all models for 5 epochs with learning rate initialized at $10^{-4}$, batch size equal to $40$ and  weight decay at $10^{-4}$. The learning rate follows a cosine annealing schedule. We use oversampling for all approaches to create balanced batches as done in \cite{bountos2021self}.
All CNN models were initialized with weights obtained from training on ImageNet. All pretrained transformer models were obtained by the model zoo of \cite{rw2019timm}. We use the base versions of all vision transformer models. Swin transformer has patch size of 4, while DeiT and ConViT have 16.

Table~\ref{tab:all_encoders_softmax_eval} presents the results of the evaluation on the C1 dataset. ACC, FP, TP, FN, TN, stand for overall accuracy, false positives, true positives, false negatives and true negatives respectively. 
The best 
accuracy is achieved by the ConvViT (87.7\%) architecture. Among CNNs the best one is achieved by the ResNet18 architecture (85\%). These results alone are encouraging, taking into account that training has been done only with synthetic data. 
Nevertheless, the classification accuracy remains lower than the current state of the art on this dataset (91\%).
This motivates us to look for more sophisticated methods that are able to overcome the covariate shift.

\begin{table}[ht]
\centering

\caption{Evaluation on the C1 dataset of standard CNNs and ViTs when trained on synthetic data. Classification 
happens using a softmax layer. ACC, FP, TP, FN, TN, stand for overall accuracy, false positives, true positives, false negatives and true negatives respectively.}
\label{tab:results}

\begin{tabular}[t]{cccccc}
    \toprule
    {Model} & {ACC} & {FP} & {TP} & {FN} & {TN}\\ 
    \midrule
    ResNet18& \textbf{85\%} & 7 & 296 & 108 & 358\\\hline
    DenseNet121 & 84.3\% & 22&306&98&343\\\hline
    VGG16 & 75.2\% & 132 & 346 & 58 &233\\\hline
    \midrule
	ConvViT & \textbf{87.7\%} & 57 & 367 & 37 & 308\\ \hline
    DeiT & 81.4\% & 115 & 376 & 28 & 250 \\ \hline
    Swin & 85.4\% & 7 & 299 & 105 & 358 \\ \hline
    \bottomrule
\end{tabular}
\label{tab:all_encoders_softmax_eval}
\end{table}

\subsection{Examining the effect of prototypes} 

In this section we investigate the effect of replacing the classic softmax classifier by a classifier which 
uses trainable prototypes as described in Section~\ref{sec:prototype-learning}. 
Prototype learning architectures 
allow us to surpass the state-of-the-art despite the fact that we only train on synthetic data.

Table~\ref{tab:eval} 
compares the performance on the C1 dataset of both CNN and Vision Transformer models when they are complemented with
prototype learning. In the case of CNNs we see that the addition of class prototypes does not provide the desired performance boost. While the performance of VGG16 slightly improves, the performance of both 
ResNet18 and DenseNet121 degrades. It is interesting that the introduction of prototypes seems to amplify 
the false negatives considerably.

On the other hand, all tested vision transformer architectures surpass the previously reported state of the art with accuracy $>91\%$. 
The boost compared to the corresponding model with a standard softmax layer 
ranges between $3.7\%$ and $11.3\%$. This enables us to create robust methods that can transfer knowledge well from the
synthetic to the real domain. It is interesting to note that the resulting models achieve their accuracy by 
balancing the errors between false positives and false negatives.

\begin{table}[ht]
    \centering
    \caption{Evaluation on the C1 dataset of the examined methods using prototype learning 
    and pseudo-labeling and the previous state of the art. 
    ACC, FP, TP, FN, TN, stand for overall accuracy, false positives, true positives,
    false negatives and true negatives respectively. }\label{tab:eval}
    
    \begin{tabular}{cccccc}
        \toprule
        {Model} & {ACC} & {FP} & {TP} & {FN} & {TN}\\ 
        \midrule
        ResNet18-PL & 78\% & 2 & 237 & 167 & 363 \\ \hline
        DenseNet121-PL & 82.4\% & 3 & 272 & 132 & 362  \\ \hline
        VGG16-PL & 82.9\% & 123 & 396 & 8 & 242 \\ \hline
        \midrule    
        ConvViT-PL &91.4\% & 28 & 366 & 38 & 337\\ \hline
        DeiT-PL & 92.7\% & 28 & 376 & 28 & 337 \\ \hline
        Swin-PL & 93.8\% & 26 & 383 & 21 & 339 \\ \hline
        \midrule     
        ConvViT-PL-Pseudo & 95.1\% & 14 & 381&23 & 351\\ \hline
        DeiT-PL-Pseudo & 93.4\% & 37 & 391  & 13 & 328\\ \hline
        Swin-PL-Pseudo & \textbf{97.1\%} & 16 & 398 & 6 & 349\\ \hline
        \midrule
        \begin{tabular}{@{}c@{}}ResNet18-Only-Pseudo\\(Section~\ref{sec:resnet-pseudo})\end{tabular}    
          & 91.6\% & 11 & 351 & 53 & 354\\ \hline
        \midrule
        ResNet50-SimCLR~\cite{bountos2021self} & 91\% & 10 & 347 & 57 & 355 \\
        \bottomrule
    \end{tabular}
    \end{table}
    
\subsection{Examination of the Domain Adaptation module}

The great boost provided by the use of learnable class prototypes motivates us to look for ways to further improve our model. To this end, we utilize an unlabeled real InSAR dataset as described in \ref{sec:pseudo}.
Table~\ref{tab:eval} also contains the evaluation of all prototype based transformer models complemented with the additional pseudo-labeled training. We use X-PL-Pseudo to denote the use of both prototype learning and the additional pseudo-training process with an encoder X. The first observation one can make is that the performance of all models increases. Our best model Swin-PL denoted as Swin-PL-Pseudo in this case achieves an impressive $97.1\%$ accuracy. Its superiority was expected given its initial $93.8\%$ accuracy on C1. The better the model that generated the pseudo-labels the less the noise it will induce in the pseudo-label training stage.

\begin{figure}[ht!]
    \begin{subfigure}{0.24\textwidth}
        \centering
       \includegraphics[width=\textwidth]{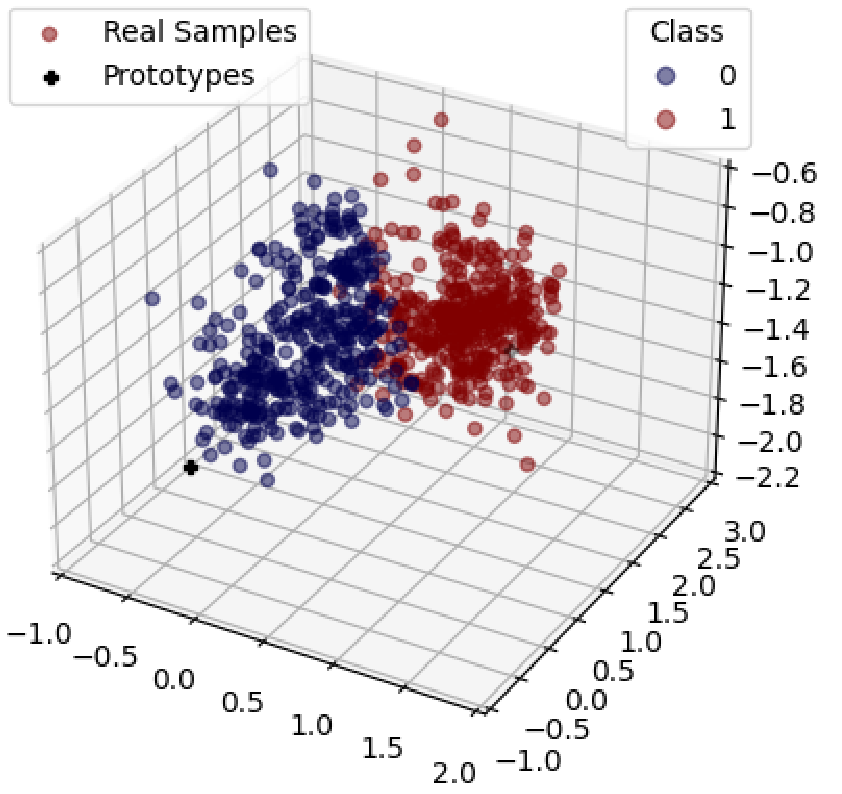}
       \caption{}
    \end{subfigure}
    \begin{subfigure}{0.24\textwidth}
    \centering
     \includegraphics[width=\textwidth]{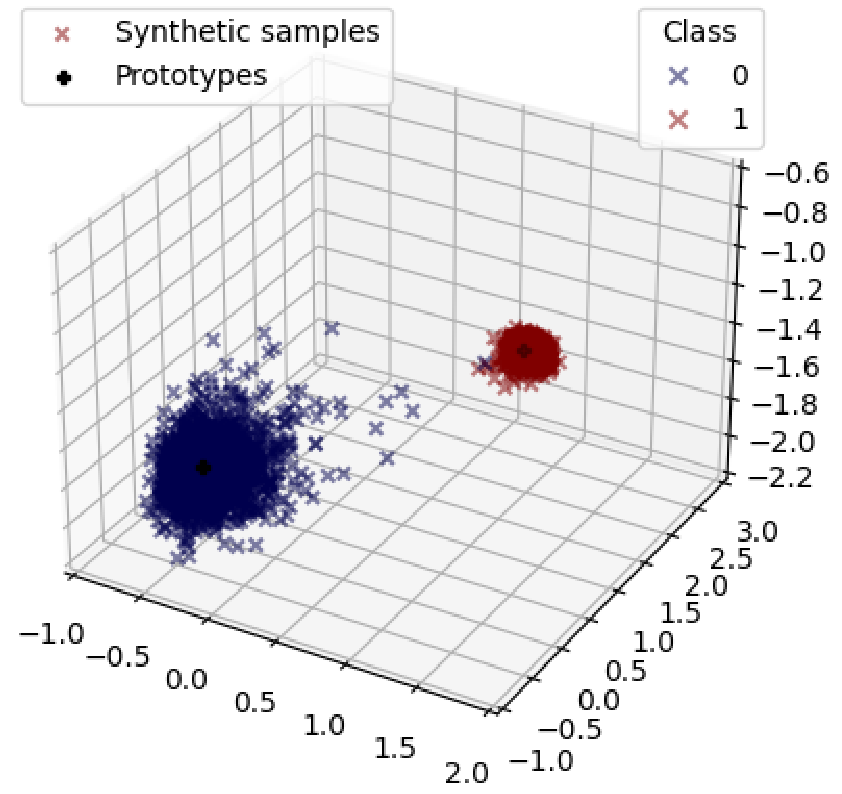}
     \caption{}
     \end{subfigure}
     \hfill
     \begin{subfigure}{0.24\textwidth}
     \centering
      \includegraphics[width=\textwidth]{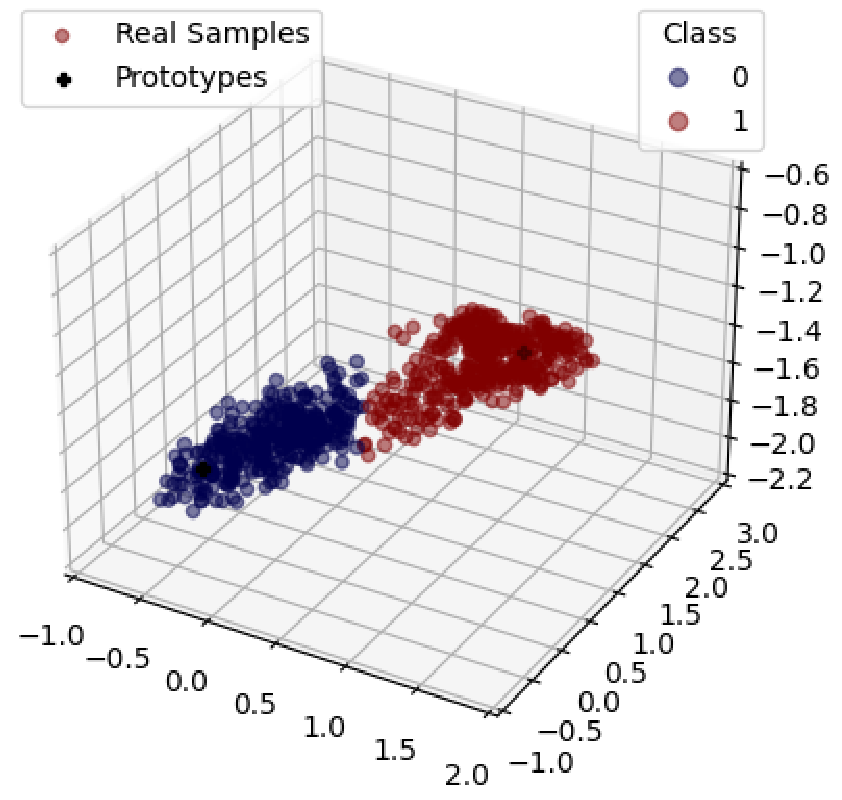}
      \caption{}
     \end{subfigure}
    \begin{subfigure}{0.24\textwidth}
    \centering
     \includegraphics[width=\textwidth]{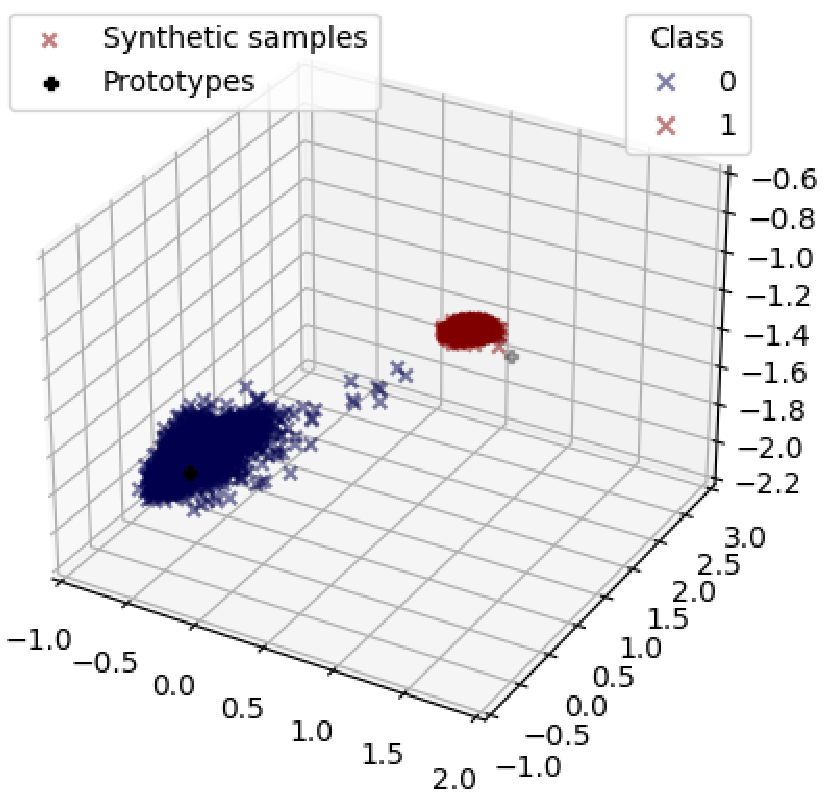}
    \caption{}
    \end{subfigure}
     \caption{Prototype Space Visualization. The top row refers to the embeddings resulting from Swin-PL while the bottom row to the respective embeddings from Swin-PL-Pseudo. Left column refers to the real domain and the right column to the synthetic domain. The major improvement lies at the class seperation on the real dataset as well as the representation compactness. On the top row the two class are blended near the border while at the bottom row the two classes are better separated.}
       \label{fig:protospace}
    \end{figure}

In order to visually examine how well the two classes are separated, we show the prototype space in Figure~\ref{fig:protospace}.
We focus on the Swin Transformer model. The top row uses Swin-PL while the bottom row Swin-PL-Pseudo. For the synthetic data there is greater inter-class distance.
This is to be expected since the model was trained on this domain. Nevertheless, after the pseudo training process described in Section~\ref{sec:pseudo}, we can see that the two classes are now more clearly separated in the real domain too. While initially the two classes were blended near their boundary, after the new non-linear projection was learnt via pseudo training the samples surround their respective prototypes and are distinctively separated from samples of the opposite class. Additionally, the performance remains stable on the synthetic domain, solidifying the conclusion on the model's ability to generalize very well.

\subsection{How does the prototype space dimensionality affect the model?}
Finally, in Table~\ref{tab:dim} we examine how the increase of the dimensionality of the prototype space affects the performance of the prototype based models. We notice that increasing the dimensions too much hurts the model massively. This could be attributed to the use of euclidean distance and the problems of distance metrics in high dimensional spaces\cite{aggarwal2001surprising}. The best performance is achieved from our base model with prototypes on a 3 dimensional space.

\begin{table}[ht]
\centering
\caption{Ablation on the effect of prototype space dimension on the C1 dataset.}\label{tab:dim}
\begin{tabular}[t]{cccccc}
    \toprule
    {Model} & {ACC} & {FP} & {TP} & {FN} & {TN}\\ 
    \midrule
    Swin-PL-3d & \textbf{93.8\%} & 26 & 383 & 21 & 339 \\ \hline 
    Swin-PL-100d& 76.46\% & 105 & 328 & 76 & 260 \\\hline
    Swin-PL-1000d & 55.5\% & 17 & 79 & 325 & 348\\\hline
    \bottomrule
\end{tabular}
\end{table}
\subsection{How do prototypes look like?}
\begin{figure}[ht]
    \centering
    \includegraphics[width=0.5\textwidth]{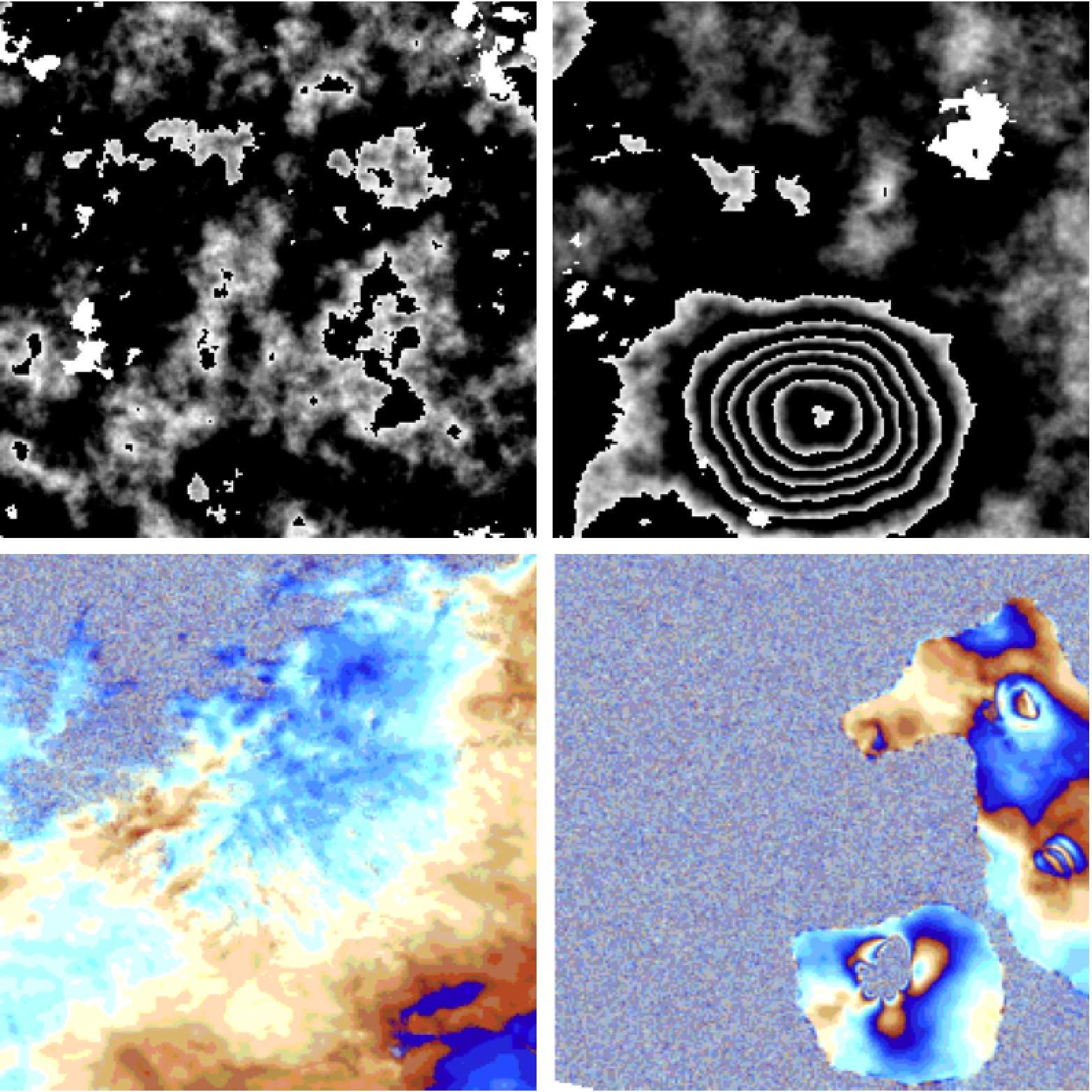}
    \caption{Visualization of the closest samples to the prototypes. On the left column, we show the negative patches and on the right the patches that
	     contain deformation. The top row contains the synthetic case where we use the standard Swin-PL architecture. Since the prototypes were learnt
	     for the synthetic domain the synthetic samples are closer to the prototypes.
	     Bottom row presents the real dataset case.
             Observing that the real samples are better projected to the prototype space after the pseudo projection training, as shown in Figure~\ref{fig:protospace}, we
	     use the Swin-PL-Pseudo model to find the closest sample.
             In both cases the prototypes remain the same.  Even though the synthetic set contains mostly larger deformation patterns, our method is able to correctly cluster
	     deformations with smaller intensity at the positive class.
}
    \label{fig:prototype_viz}
\end{figure}

Learning class prototypes has another obvious benefit besides the improved accuracy. It gives us a direct way to investigate how our model discriminates between the two classes by examining the class prototypes leading to more explainable classifiers. However, since our model is not invertible, we cannot directly generate an input sample based on the final 3-d projection in the prototype space. In order to visually inspect how the class prototypes look like we investigate their closest samples. We do that for both real and synthetic domains. Obviously, since the prototypes were found while training on the synthetic data, these samples will be closer and constitute a more realistic representation. However, we find the results from the real domain quite interesting and present them along with the respective figures from the synthetic domain. The prototypes visualization can be seen in Figure~\ref{fig:prototype_viz}.

\subsection{How do positional embeddings look like in vision transformers?}
\begin{figure}
\includegraphics[width=0.5\textwidth]{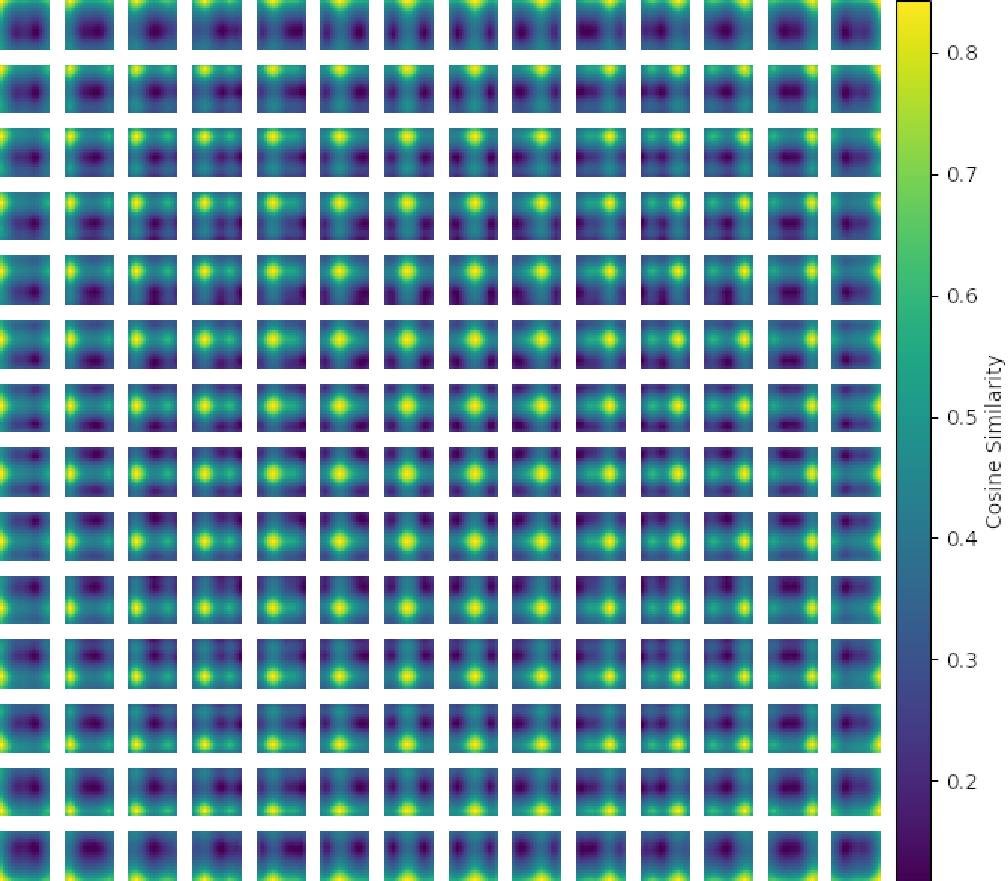}
\caption{Cosine similarity of positional embeddings among patches in DeiT. Each mini patch shows the cosine similarity between its position embedding and the embeddings of the rest of the patches in different positions. The full grid represents the whole image split into 14x14 patches with patch size equal to 16.}
\label{fig:positional_embeddings}
\end{figure}

Swin Transformers directly induce locality information in vision transformer architectures. DeiT however, learns positional information from scratch. DeiT had the greatest classification accuracy boost from the use of class prototypes and the second highest accuracy in the simple prototype learning setting. It is worthy to examine what kind of information do the positional embeddings provide and how do they differ between different locations in the image. Figure \ref{fig:positional_embeddings} shows the full 14x14 grid of patches along with their cosine similarity with the rest of the patches. Each patch location is depicted with a mini patch showing the cosine similarity of the specific location in comparison to the rest. It is clear that certain patterns are inferred by the network. The high cosine similarity between close points and low between points that are further away shows that the learnt embeddings encode a notion of distance. Additionally, elements from the same row and column have increased similarity. Dosovitskiy et~al.~\cite{dosovitskiy2020image} reached to the same conclusions. 

\subsection{Can we extract meaningful insights about the model's decision from the learnt self-attention?}
\begin{figure}
    \centering
    \begin{subfigure}{0.24\textwidth}
            \includegraphics[width=\textwidth]{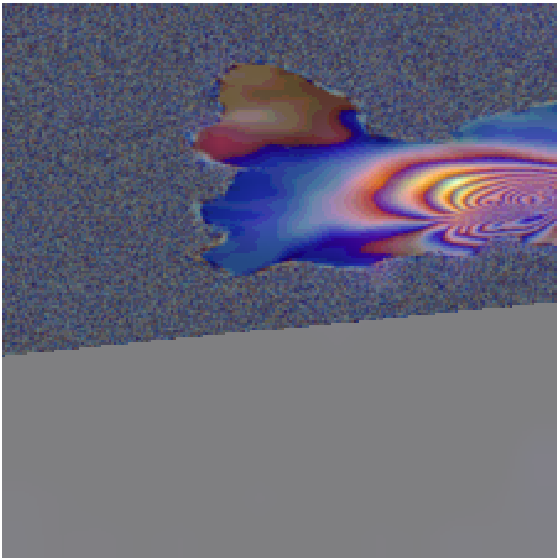}
    \end{subfigure}
    \begin{subfigure}{0.24\textwidth}
        \includegraphics[width=\textwidth]{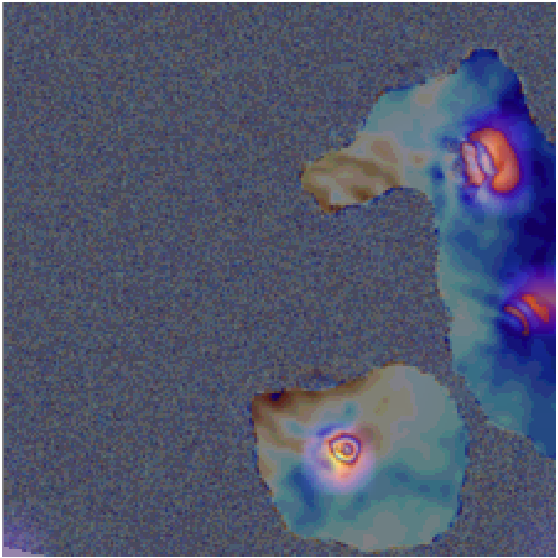}
    \end{subfigure}
    %\hfill
    %\vspace{1cm}
    \par\smallskip
    \begin{subfigure}{0.24\textwidth}
        \includegraphics[width=\textwidth]{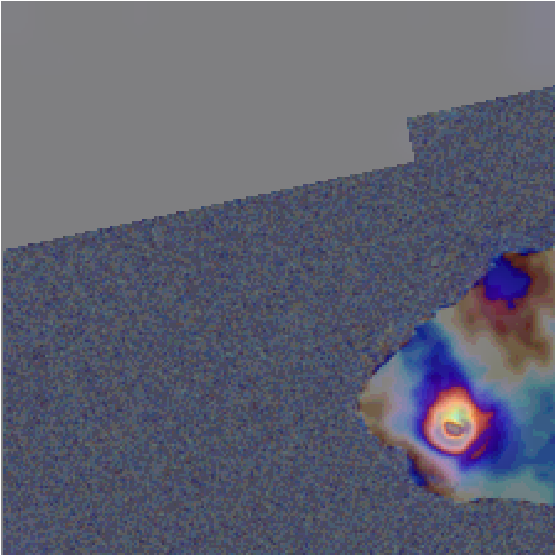}
    \end{subfigure}
    \begin{subfigure}{0.24\textwidth}
        \includegraphics[width=\textwidth]{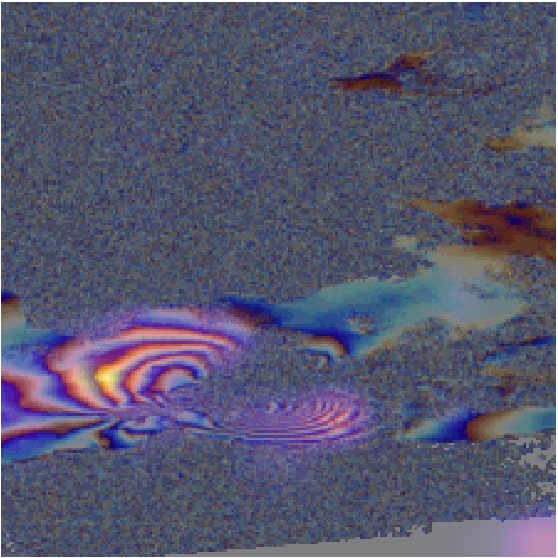}
    \end{subfigure}
    \caption{Attention based interpretation of the focus of DeiT-PL's decisions.}
    \label{fig:viz_attention}
\end{figure}

In this section we investigate the learnt attention of our top models. We start with Deit-PL by using a gradient attention rollout technique to create class specific visualizations. Figure~\ref{fig:viz_attention} shows the respective plots. We can see excellent localization focus to the deformation fringes
from DeiT-PL. 

\begin{figure}
    \centering
            \begin{subfigure}{0.15\textwidth}
           \includegraphics[width=\textwidth]{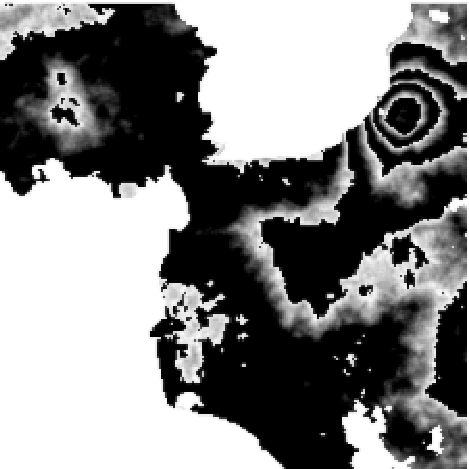}
           \caption{}
                \end{subfigure}
        \begin{subfigure}{0.15\textwidth}
           \includegraphics[width=\textwidth]{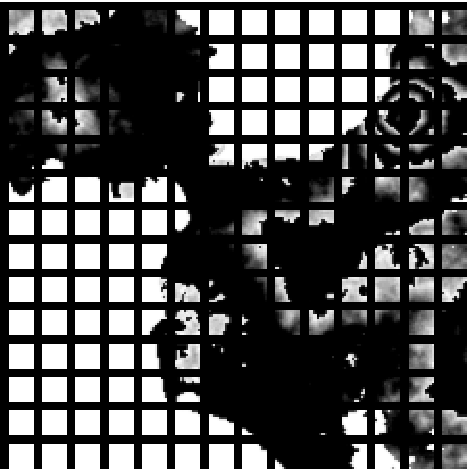}
           \caption{}
                \end{subfigure}
        \begin{subfigure}{0.15\textwidth}
           \includegraphics[width=\textwidth]{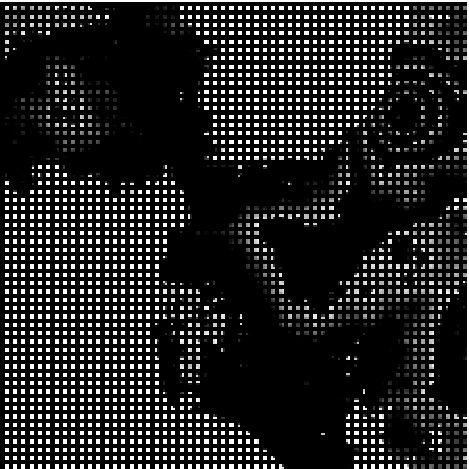}
           \caption{}
        \end{subfigure}
     \caption{Input sample for self-attention visualization. a) shows the original sample while b) and c) show the respective split in 16x16 and 4x4 patches by DeiT and Swin Transformer respectively.}     
    \label{fig:input_sample}
\end{figure}

\begin{figure}[ht!]
    \centering
    \begin{subfigure}{0.24\textwidth}
    \includegraphics[width=\textwidth]{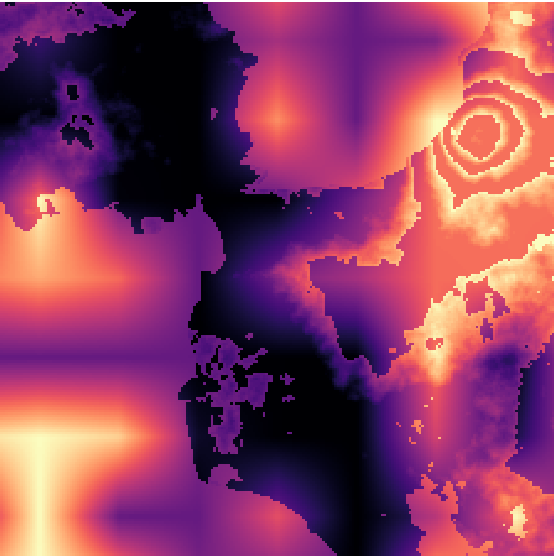}
    \end{subfigure}
    \begin{subfigure}{0.24\textwidth}
        \includegraphics[width=\textwidth]{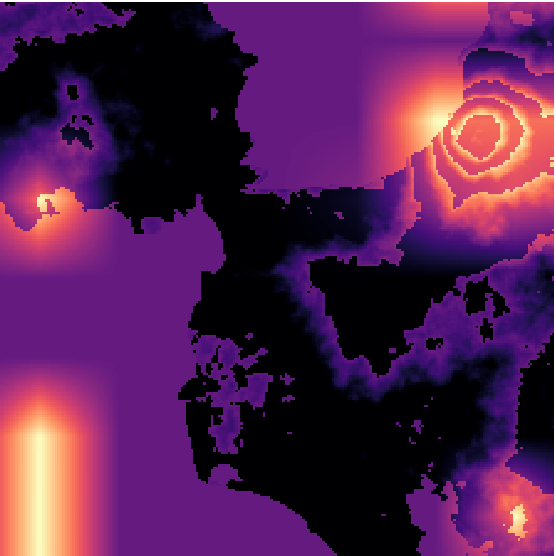}
    \end{subfigure}
    \hfill
    \par\smallskip
    \begin{subfigure}{0.24\textwidth}
        \includegraphics[width=\textwidth]{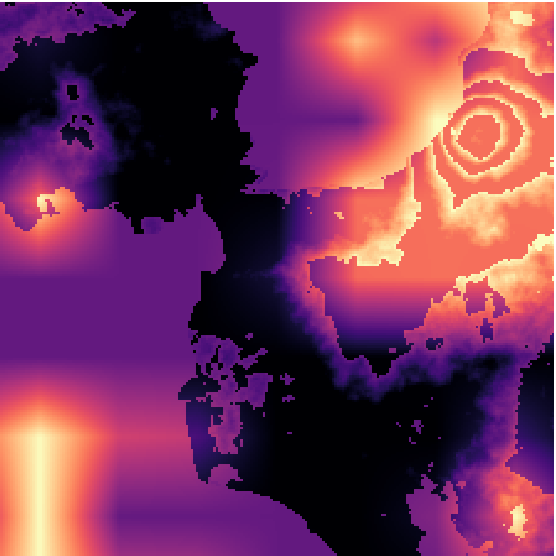}
    \end{subfigure}
    \begin{subfigure}{0.24\textwidth}
        \includegraphics[width=\textwidth]{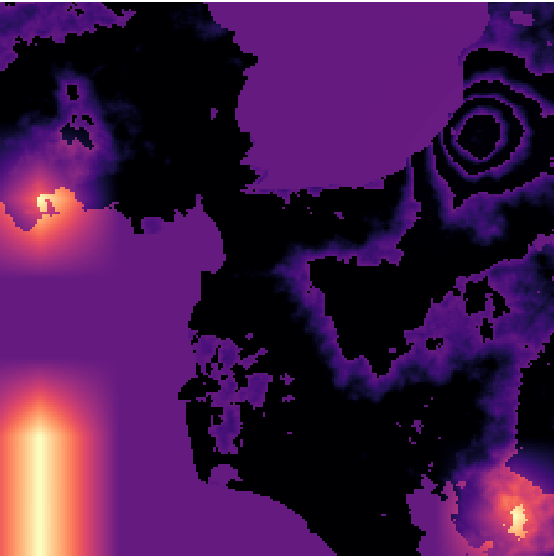}
    \end{subfigure}
    \caption{Visualization of the attention maps of the first four attention heads of Swin-PL transformer encoder's last layer.}
    \label{fig:swin_attention}
\end{figure}

In the case of Swin-PL we directly explore the self-attention map of the last layer of the Swin Transformer.
We focus on the middle pixel of the window and visualize where it attends the most.  
Swin Transformer attention can be seen in Figure~\ref{fig:swin_attention} using the patch of Figure~\ref{fig:input_sample} as input.
We see the first four attention heads for the self-attention of the middle pixel, and observe that in 3 out of 4 heads the deformation fringe has high attention. Second head (up-right) in particular, focuses more on
the core of the deformation. Each head learns different things and their combination leads to good model performance.
As stated in~\cite{chefer2021transformer}, using solely the attention
map to explain the model's reasoning is naive, since there are a lot more layers and processes in a model that contribute to its
decisions. Nevertheless, even with this limited view, we can see the good perception of the model on what constitutes a
ground deformation pattern.

\subsection{Can we generate a new trustworthy pseudo labeled InSAR dataset using our models?}\label{sec:resnet-pseudo}

In this subsection we discuss whether we can use synthetic data to create new InSAR datasets on the real domain. For this 
purpose we utilize the great performance of our best model Swin-PL-Pseudo on the real test set (97.1\%).
Looking at the breakdown of Table~\ref{tab:eval} we see that the model produced 16 false positives out of 365 negatives and
6 false negatives out of 404 positives. The low number of false negatives is important for a critical task like that.

Given the above observations, we used Swin-PL-Pseudo to create pseudo labels on a new unlabeled real InSAR dataset with 2272, $224\times224$ pixel samples, acquired from Comet-LiCS portal, the same source as C1.
We then use these pseudo labels to train a simple ResNet18 CNN for 100 epochs, keeping the training settings as described in
Section~\ref{sec:best-encoder}, while removing the weight decay since the induced noise of the pseudo-labels can act as a regularizer. The test accuracy of the resulting model ($91.6\%$) on C1 surpasses the previous state of the
art ($91\%$) and by far the performance of the ResNet18 model when trained on synthetic data ($85\%)$. 
We denote this model as ResNet18-Only-Pseudo in Table~\ref{tab:eval}.

\begin{figure*}[ht]
\centering
\includegraphics[width=\textwidth]{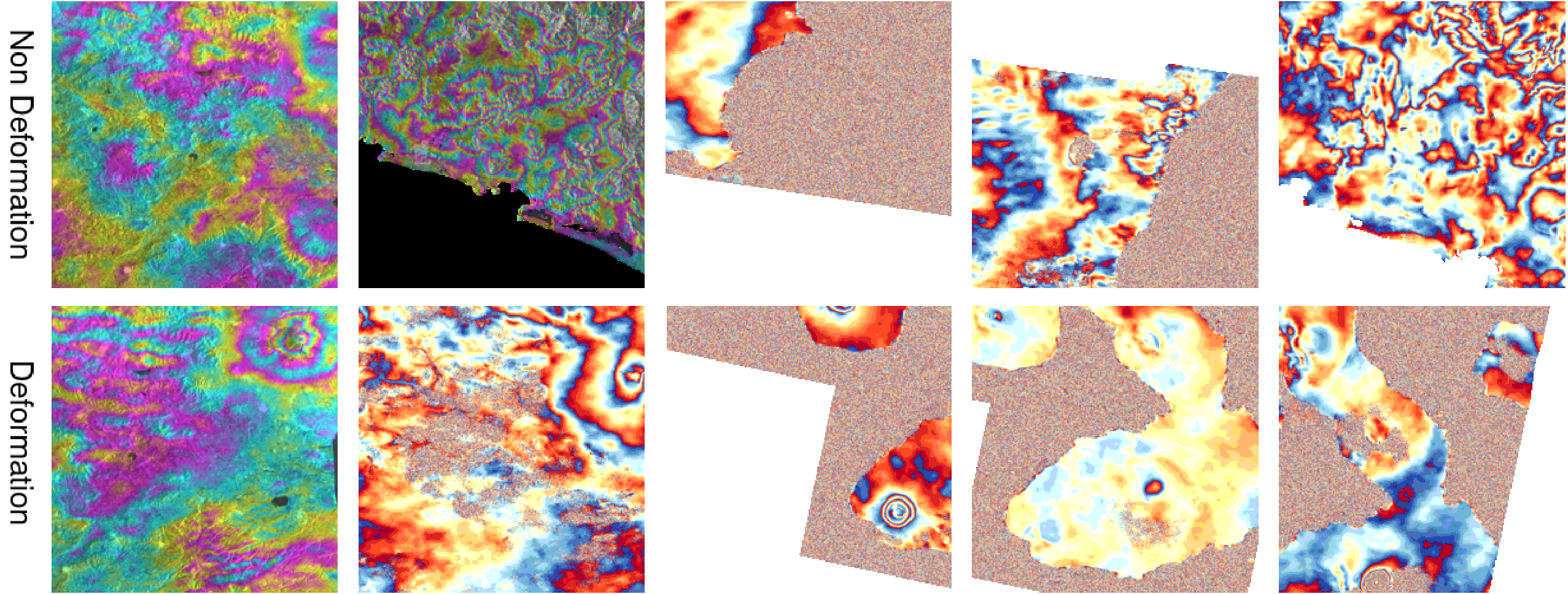}
\caption{Qualitative examination of pseudo labels assigned to the unlabeled dataset using the Swin-PL-Pseudo model. These labels were used to train a ResNet18 with more than 6\% higher accuracy in comparison to the respective model trained on synthetic data. We provide 5 challenging samples from each assigned category. The top row shows samples predicted as non-deformation while the bottom shows patches labeled as deformation.}
\label{fig:pseudo_labeled_data}

\end{figure*}

The $91.6\%$ accuracy shows that it learnt quality features in a supervised setting. If the supervision was faulty, with high
levels of noise, its accuracy would not be that high, and definitely not higher than the previous state of the art on C1 from
\cite{bountos2021self}. However, quantitative evaluation of such a task is not enough. Figure~\ref{fig:pseudo_labeled_data}
presents a few challenging samples along with the pseudo label produced by Swin-PL-Pseudo. The top row contains non-deformation samples while the bottom row patches labeled as deformation. In the second and fifth column of the first row, we see patches containing multiple fringes. These fringes do not follow the volcanic deformation spatial pattern, are most likely caused by atmospheric disturbances and were correctly classified as non-deformation. The most distinctively encouraging results however, lie on the bottom row. In the first column potential deformation fringes are visible. However, an expert would require additional information to classify this interferogram, such as a digital elevation model, since such fringes could be the expression of a topography related atmospheric phase screen. This applies for all positive pseudo labels. In the second column, the deformation on the top right is well identified despite the surrounding noise. The third and fourth columns show impressive results. Both patches contain small deformation fringes, especially in the fourth column. Fringes like that are non-existent in the synthetic dataset. Furthermore, in the final example we observe fringes hidden behind noise. This noise could be attributed to the existence of water or other SAR signal decorrelation factors that affect interferometric coherence. The ability to identify the existence of deformation under these conditions is of utmost importance in this critical task. 
Hence, considering the great range of applications of InSAR, and the technology that can provide us with synthetic data, our proposed method can create quality labeled datasets that will require minimum to no human supervision and still be able to train simple models for the task at hand.

\section{Conclusion}

In this work we systematically tackle the problem of volcanic unrest detection using synthetically generated InSAR data. Our 
framework, based on prototype learning and vision transformers, generalizes to the real domain and surpasses the current state of
the art ($>93\%$) on the C1 test set. 
Additionally, given an unlabeled dataset from the real-domain, we utilize self-labeling and boost the performance of our model up
to $97.1\%$ without geopardizing its generalization abilities, retaining its strong classification performance on the synthetic validation set. 
To showcase the robustness of our model, we use it to pseudo-label a real dataset and train a simple ResNet18 that achieves
$91.6\%$ on our test set, higher than the same architecture when trained using synthetic data. 
Finally, we explore the nature of the learnt prototypes as well as the properties of our encoder, including its self-attention and its positional embeddings, towards a more interpretable method for volcanic unrest detection. 

Due to the general nature of our methodology and the combined strength of prototype learning and vision transformers, the approach 
should be applicable to other downstream tasks with InSAR data such as localisation of deformation, and detangling of atmospheric, orbital, Digital Elevation Model and deformation interferometric signals.

\section*{Acknowledgment}
This work has received funding from the European Union’s Horizon2020 research and innovation project DeepCube, under grant agreement number 101004188. LiCSAR contains modified Copernicus Sentinel data [2014-2021] analysed by the Centre for the Observation and Modelling of Earthquakes, Volcanoes and Tectonics (COMET). LiCSAR uses JASMIN, the UK’s collaborative data analysis environment (http://jasmin.ac.uk).

\end{document}